\documentclass[preprints,article,accept,moreauthors,pdftex]{Definitions/mdpi} 

\firstpage{1} 
\pubvolume{1}
\issuenum{1}
\articlenumber{0}
\pubyear{2021}
\copyrightyear{2020}
\datereceived{} 
\dateaccepted{} 
\datepublished{} 
\hreflink{https://doi.org/} 

\usepackage{array}
\newcolumntype{C}[1]{>{\centering\arraybackslash}m{#1}}

\usepackage{xcolor}
\usepackage{colortbl}

\newdimen\NetTableWidth
 \NetTableWidth=\dimexpr
    \linewidth - 8\tabcolsep - 5\arrayrulewidth
\usepackage{multirow}

\usepackage{soul}
\soulregister\ref7
\soulregister\cite7

\usepackage{float}
\usepackage{stfloats}

\usepackage{subcaption}

\usepackage{algorithm}
\usepackage{algpseudocode}

\Title{Utilising Flow Aggregation to Classify Benign Imitating Attacks}

\TitleCitation{Utilising Flow Aggregation to Classify Benign Imitating Attacks}

\Author{
Hanan Hindy $^{1,}$\orcidA{}, 
Robert Atkinson $^{2}$\orcidB{}, 
Christos Tachtatzis $^{2}$\orcidC{}, 
Ethan Bayne $^{1}$\orcidD{},
Miroslav Bures $^{3}$\orcidE{},
and Xavier Bellekens $^{2,}$*\orcidF{}}

\AuthorNames{Hanan Hindy, Robert Atkinson, Christos Tachtatzis, Ethan Bayne, Miroslav Bures and Xavier Bellekens}

\AuthorCitation{Hindy, H.; Atkinson, R.; Tachtatzis, C.; Bayne, E.; Bures, M.; Bellekens X.}

\address{%
$^{1}$ \quad Division of Cybersecurity, Abertay University, Dundee DD1 1HG, UK; hananhindy@ieee.org, e.bayne@abertay.ac.uk\\
$^{2}$ \quad Electronic and Electrical Engineering Department, University of Strathclyde, Glasgow G1 1XQ, UK; \{robert.atkinson,christos.tachtatzis,xavier.bellekens\}@strath.ac.uk \\
$^{3}$ \quad Department of Computer Science, FEE, Czech Technical University, Prague, Czechias; buresm3@fel.cvut.cz}

\corres{Correspondence: xavier.bellekens@strath.ac.uk (X.B.)}

\abstract{Cyber-attacks continue to grow, both in terms of volume and sophistication. This is aided by an increase in available computational power, expanding attack surfaces, and advancements in the human understanding of how to make attacks undetectable. Unsurprisingly, machine learning is utilised to defend against these attacks. 
In many applications, the choice of features is more important than the choice of model. A range of studies have, with varying degrees of success, attempted to discriminate between benign traffic and well-known cyber-attacks. The features used in these studies are broadly similar and have demonstrated their effectiveness in situations where cyber-attacks do not imitate benign behaviour. To overcome this barrier, in this manuscript, we introduce new features based on a higher level of abstraction of network traffic. Specifically, we perform flow aggregation by grouping flows with similarities. This additional level of feature abstraction benefits from cumulative information, thus qualifying the models to classify cyber-attacks that mimic benign traffic.
The performance of the new features is evaluated using the benchmark CICIDS2017 dataset and the results demonstrate their  validity and effectiveness.
This novel proposal will improve the detection accuracy of cyber-attacks and also, build towards a new direction of feature extraction for complex ones. }


\begin{document}
\section{Introduction}
\label{sec:introduction}
Internet traffic pattern analysis is an established and mature discipline. A key application of this research domain is the detection of malicious network intrusions. Intrusion Detection Systems~(IDS) research started in the late 1980s. Statistical models were the first to be introduced. Later, signature-based IDSs were proposed~\cite{RefWorks:doc:5f8d6428e4b0bd986d35b768}. Their detection depended on known patterns (signatures) that were used to distinguish between malicious and benign traffic flows. More recently, the volume and sophistication of these attacks has increased and that has motivated the use of machine learning~(ML) techniques to counteract them~\cite{hindy2020leveraging}. 

In many machine learning applications, it is well known that the choice of features, i.e., the inputs fed into the model, is more important than the choice of the model~\cite{RefWorks:doc:5cdd83f9e4b09410f5d04260}. Indeed, Ghaffarian and Shahriari state that features play a vital role in the development of IDS~\cite{RefWorks:doc:5ae8746ae4b029aa3efa7c2b}. When Internet traffic is analysed on a flow-by-flow basis, the choice of features is quite self-evident: are particular flags set or reset in the internet packet headers, what is the average size of a packet in a flow, what is the standard deviation of packet sizes, what is the average time between packets in a flow, etc. In simple terms, values are extracted from packet headers and various statistics are extracted from the packet lengths and inter-arrival times~\cite{alaidaros2011overview}. These features have proved adequate for identifying previous generations of cyber-attacks using a range of ML models; however, they have proven to be inadequate to the latest most sophisticated attacks.

To address this problem of detecting complex cyber-attacks, specifically the benign imitating ones, we propose, in this paper, an additional level of feature abstraction, named `\textit{Flow Aggregation}'. With this approach, we look at flows at a higher level of abstraction by bundling similar flows and extracting features across them. This form of aggregation permits a deeper representation of network traffic, and increases the performance of classification models, particularly for the more sophisticated attacks that attempt to closely resemble benign network traffic and hence evade detection. The proposed features are evaluated using the benchmark CICIDS2017~\cite{RefWorks:doc:5d4acfdae4b002b95f900284} dataset with a particular focus on the attacks that have proven most difficult to detect using well-adopted features.

The contributions of this paper are:
\begin{itemize}
    \item The introduction of a higher level of abstraction for network traffic analysis by proposing novel features to describe bundles of flows.
    \item Performance improvements in binary classification of cyber-attacks when these novel features are utilised, particularly for attacks that mimic benign network traffic. 
    \item Performance improvements in multi-class classification of cyber-attacks when these novel features are utilised.
    \item Performance improvements in zero-day attack detection when these novel features are utilised.
\end{itemize}

The remainder of the paper is organised as follows; Section~\ref{sec:relatedwork} outlines the related work and the required background. The methodology is explained in Section~\ref{sec:methodology}. Section~\ref{sec:experiments} discusses the proposed feature abstraction and the conduction of experiments alongside the results. Finally, Section~\ref{sec:conclusion} concludes the paper. 
\section{Background and Related Work}
\label{sec:relatedwork}

As mentioned, the detection of cyber-attacks is an established research area that has leveraged a range of technologies as it has evolved over the years~\cite{RefWorks:doc:5f8d65bbe4b0ca7b275231a2} to cope with the exponential growth of cyber-attacks~\cite{sarker2020cybersecurity}. A range of ML-based models have been applied to the problem, including support vector machines, artificial neural networks, and k-means clustering~\cite{9108270}. Despite the discriminative power of these models, many cyber-attacks still remain undetected or have low rates of detection.


Older, more well-recognised, attacks have been captured in KDD Cup'99 and the NSL-KDD datasets. These can be used to train ML-based models, and in many cases achieve good results. More contemporary cyber-attacks have been recorded in the CICIDS2017 dataset~\cite{RefWorks:doc:5b227bb8e4b07f83f15ddb45}. Much of the research involving this dataset has considered a subset of attacks that are particularly distinctive from benign (i.e., DDoS, Port Scan, and Web attacks), and good results have been achieved. However, some classes are either (a)~left undetected, due to their similar behaviour to normal traffic, thus difficult to detect, or (b)~when used in detection models, their low detection accuracy is concealed in the overall accuracy due to the class imbalance problem of this dataset~\cite{panigrahi2018detailed}. In this paper, we focus on those attacks that have thus far proven elusive for researchers to identify reliably, \emph{viz.} DoS Slowloris and DoS SlowHTTPTest.

Many studies use performance metrics that do not take into consideration the imbalance between relatively few examples of cyber-attacks and overwhelming examples of benign traffic. For example, an `always zero' classifier (that always predicts benign traffic) would appear to be 99\% accurate if the dataset it was tested on had 99 examples of benign traffic to each example of cyber-attack traffic. Clearly, this has the potential to provide grossly misleading results since it would never detect a single attack. To address this shortcoming, in this paper, we will fully disclose the results for precision, recall, and F1-score for each class independently.


Table~\ref{tab:recent-papers-1} and Table~\ref{tab:recent-papers-2} provide a comprehensive list of recent studies in which \mbox{CICIDS2017} dataset has been used. The tables present the published papers, the models/techniques applied, the metrics utilised to assess performance, and the concomitant results. By observing Table~\ref{tab:recent-papers-1}, two findings are highlighted. Firstly, some classes (SSH, DDoS, and Port Scan, for example) have received significant attention from researchers, whilst others have been largely ignored due to their poor results {and their benign-like behaviour that makes their classification difficult}.  Secondly,  the overall accuracy is much higher than the accuracy for individual classes. For example, in~\cite{8681044}, when the authors use 1-layer Deep Neural Network~(DNN), the overall multi-class classification accuracy in is 96\% (Table~\ref{tab:recent-papers-1}) while the individual classes detection accuracies are 55.9\%, 95.9\%, 85.4\% and 85.2\% for normal, SSH, DDoS and port scan classes, respectively (Table~\ref{tab:recent-papers-1}). Similar behaviour when the authors use 5-layers DNN as demonstrated in Table~\ref{tab:recent-papers-1}. This indicates the misleading effect of reporting the overall accuracy when dealing with imbalanced datasets. 

\begin{specialtable}[H]
    \centering
    \caption{CICIDS2017 Recent Papers Performance Summary (1). \label{tab:recent-papers-1}}
    \small
    \begin{tabular}{C{0.1\NetTableWidth}C{0.1\NetTableWidth}C{0.15\NetTableWidth}C{0.15\NetTableWidth}C{0.125\NetTableWidth}C{0.125\NetTableWidth}C{0.125\NetTableWidth}C{0.125\NetTableWidth}}
    \toprule
    \textbf{Year} & \textbf{Ref} & \textbf{Approach} & \textbf{Covered Attacks} & \textbf{Accuracy} & \textbf{Precision} & \textbf{Recall} & \textbf{F-Score} \\ \midrule
    \multirow{4}{*}{2020} & \multirow{4}{*}{\cite{9118459}\textsuperscript{+}} & MLP & \multirow{2}{*}{SSH} & - & 82\% & 98\% & 90\% \\ \cline{3-3} \cline{5-8}
    
    & &  LSTM & &  - & 97\% & 98\% & 97\% 
    \\ \cline{3-8}
    
    & &  MLP & \multirow{2}{*}{FTP} & - & 93\% & 77\% & 85\%
    \\ \cline{3-3} \cline{5-8}
    
    & &  LSTM & & - & 98\% & 99\% & 99\% 
    \\ \hline

    \multirow{12}{*}{2019} & \multirow{12}{*}{\cite{8681044}\textsuperscript{+}} & DNN (1 Layer) & \multirow{6}{*}{Binary} & 96.3\% & 90.8\% & 97.3\% & 93.9\% 
    \\ \cline{3-3} \cline{5-8}
    
    &  & DNN (5 Layers) &  & 93.1\% & 82.7\% & 97.4\% & 89.4\%  
    \\ \cline{3-3} \cline{5-8}
     
    & & LR & & 83.9\% & 68.5\% & 85\% & 75.8\%  
    \\ \cline{3-3} \cline{5-8}
        
    & & NB & & 31.3\% & 30\% & 97.9\% & 45.9\%  
    \\ \cline{3-3} \cline{5-8}
    
    & & KNN & & 91.0\% & 78.1\% & 96.8\% & 86.5\%  
    \\ \cline{3-3} \cline{5-8}
    
    & & SVM (RBF) & & 79.9\% & 99.2\% & 32.8\% & 49.3\%  
    \\ \cline{3-8} 
    
     &  & DNN (1 Layer) & \multirow{6}{*}{Multi-class} &   96\% & 96.9\% & 96\% & 96.2\% 
    \\ \cline{3-3} \cline{5-8}
      
    &  & DNN (5 Layers) &  &  95.6\% & 96.2\% & 95.6\% & 95.7\% 
    \\ \cline{3-3} \cline{5-8}
    
    & & LR & &   87\% & 88.9\% & 87\% & 86.8\% 
    \\ \cline{3-3} \cline{5-8}
    
    & & NB & &  25\% & 76.7\% & 25\% & 18.8\%  
    \\ \cline{3-3} \cline{5-8}
    
    & & KNN & &   90.9\% & 94.9\% & 90.9\% & 92.2\% 
    \\ \cline{3-3} \cline{5-8}
    
    & & SVM (RBF) & &   79.9\% & 75.7\% & 79.9\% & 72.3\% 
    \\ \hline
    
    2019 & \cite{yulianto2019improving} & AdaBoost & DDoS &  81.83\% & 81.83\% & 100\% & 90.01\% \\ \hline
    
    \multirow{2}{*}{2018} & \multirow{2}{*}{\cite{8625370}} & DL & \multirow{2}{*}{PortScan} & 97.80\% & 99\% & 99\% & 99\% \\ \cline{3-3} \cline{5-8}
    
    & & SVM & &  69.79\% & 80\%  & 70\% & 65\%\\ \hline
    
    \multirow{4}{*}{2018} & \multirow{4}{*}{\cite{abdulrahman2018evaluation}} & C5.0 & \multirow{4}{*}{DDoS} & 85.92\%  & 86.45\% & 99.70\% & -
    \\ \cline{3-3} \cline{5-8}
    
     & & RF & & 86.29\% &  86.80\% &  99.63\% & -
     \\ \cline{3-3} \cline{5-8}
     
     & & NB & & 90.06\% & 79.99\% &  86.03\% & -
     \\ \cline{3-3} \cline{5-8}
     
     & & SVM & & 92.44\% & 79.88\% & 84.36 & -
     \\ \bottomrule
     
     \multicolumn{8}{l}{\textsuperscript{+}: Only snippets of the results are listed in the table.} \\
     \multicolumn{1}{l}{Where:} & \multicolumn{3}{l}{DDoS: Distributed Denial of Service} & \multicolumn{4}{l}{MLP: Multilayer Perceptron} \\
     \multicolumn{1}{l}{} & \multicolumn{3}{l}{DL: Deep Learning} & \multicolumn{4}{l}{NB: Na\"ve Bayes} \\
     \multicolumn{1}{l}{} & \multicolumn{3}{l}{DNN: Deep Neural Network} & \multicolumn{4}{l}{RBF: Radial Basis Function} \\
     \multicolumn{1}{l}{} & \multicolumn{3}{l}{FTP: File Transfer Protocol} & \multicolumn{4}{l}{RF: Random Forest} \\ 
     \multicolumn{1}{l}{} & \multicolumn{3}{l}{KNN: k-Nearest Neighbour} & \multicolumn{4}{l}{SSH: Secure Shell} \\
     \multicolumn{1}{l}{} & \multicolumn{3}{l}{LR: Logistic Regression} & \multicolumn{4}{l}{SVM: Support Vector Machine} \\
     \multicolumn{1}{l}{} & \multicolumn{3}{l}{LSTM: Long short-term memory} & \multicolumn{4}{l}{} \\
     
    \end{tabular}
\end{specialtable}

\begin{specialtable}[H]
    \centering
    \caption{CICIDS2017 Recent Papers Performance Summary (2). \label{tab:recent-papers-2}}
    \small
    \begin{tabular}{C{0.1\NetTableWidth}C{0.1\NetTableWidth}C{0.15\NetTableWidth}C{0.125\NetTableWidth}C{0.125\NetTableWidth}C{0.125\NetTableWidth}C{0.125\NetTableWidth}}
    \toprule
     & &  & \multicolumn{4}{c}{\textbf{Accuracy}} \\ \cline{4-7}
     \multirow{-2}{*}{\textbf{Year}} & \multirow{-2}{*}{\textbf{Ref}} & \multirow{-2}{*}{\textbf{Approach}} & \textbf{Normal} & \textbf{SSH} & \textbf{DDoS} & \textbf{PortScan} \\ \midrule

    \multirow{6}{*}{2019} & \multirow{6}{*}{\cite{8681044}\textsuperscript{+}} 
      & DNN (1 Layer)  & 55.9\% & 95.9\% & 85.4\%  & 85.2\% \\  \cline{3-7}
    
    & & DNN (5 Layers)  & 56.8\% & 95.8\% & 85.5\% & 85.5\%  \\ \cline{3-7}

    & & LR & 88.5\% & 98.4\% & 92.2\% & 92.6\% \\ \cline{3-7}
    
    & & NB  & 32.2\% & 75.7\% & 98.5\%  & 87.9\%  \\ \cline{3-7}
    
    & & KNN & 90.9\% & 97\% & 99.5\% & 99.6\% \\ \cline{3-7}
    
    & & SVM (RBF) &  79.8\% & 98.8\% & 92.9\% & 99\%   \\ \bottomrule

     \multicolumn{7}{l}{\textsuperscript{+}: Only snippets of the results are listed in the table.} \\
    \end{tabular}
\end{specialtable}

Furthermore, Vinayakumar~\textit{et al.}~\cite{8681044} highlighted in their recent research that by observing the saliency map for the CICIDS2017 dataset, it is shown that ``the dataset requires few more additional features to classify the connection record correctly''~\cite{8681044}. The authors' observations highlighted this need for the Denial of Service~(DoS) class specifically. As later discussed in Section~\ref{sec:experiments}, this concurs with our findings regarding the attack classes that require the additional abstraction level of features to be differentiated from benign traffic and other attacks.

\subsection{Feature Engineering}

In the ML domain, features are used to represent a measurable value, a characteristic or an observed phenomenon~\cite{bishop2006pattern}. Features are informative: they are usually represented numerically, however, some can be in categorical or string format. When categorical features are used for ML, encoding is used to transform them into a numerical (ML-friendly) format. 

Obtaining features can be done by construction, extraction or selection processes, or a combination of them~\cite{RefWorks:doc:5f8d6655e4b00c14b3f5d5e6}.
\textbf{Feature construction} creates new features by mining existing ones by finding missing relations within features. 
While \textbf{extraction} works on raw data and/or features and apply mapping functions to extract new ones. 
\textbf{Selection} works on getting a significant subset of features. This helps reduce the feature space and reduce the computational power. 
Feature selection can be done through three approaches, as shown in Table~\ref{tab:filter-selection}; filter, wrapper, and embedded.

\begin{specialtable}[H]
    \caption{Feature Selection Approaches. \label{tab:filter-selection}}
    \begin{center}
    \small
    \begin{tabular}{C{0.25\NetTableWidth}C{0.25\NetTableWidth}C{0.25\NetTableWidth}C{0.25\NetTableWidth}}
    \toprule
        \textbf{Approach} & \textbf{Description} & \textbf{Advantages} & \textbf{Disadvantages}\\
    \midrule
        Filter~\cite{hamon:tel-00920205} & Selects the most meaningful features regardless of the model & Low Execution Time and over-fitting & May choose redundant variables \\
    \hline
        Wrapper~\cite{phuong2005choosing} & Combine related variables to have subsets & Consider interactions & Over-fitting risk and High execution time  \\
    \hline
        Embedded~\cite{hernandez2007genetic} & Investigate interaction more thoroughly  than Wrapper & Result in an optimal subset of variables & -- \\  
    \bottomrule
    \end{tabular}
    \end{center}
    
\end{specialtable}

Law~\textit{et~al.}~\cite{1316850} highlight the importance of feature selection for ML models. The authors discuss its effect on boosting performance and reducing the effect of noisy features, specifically when training using small datasets. Furthermore, nonuniform class distributions should be considered to avoid misleading results when applying a supervised feature selection~\cite{PUDIL19951389}.
Alternatively, when the dataset labels are not available, an unsupervised feature selection is used. Mitra~\textit{et~al.}~\cite{990133} categorise unsupervised feature selection techniques into (a)~clustering performance maximisation and (b)~feature dependencies and relevance. 

\subsection{Artificial Neural Network}
ANNs are inspired by the human biological brain. McCulloch and Pitts~\cite{mcculloch1943logical} proposed the first ANN in 1943. Later in 1986, Rumelhart and McClelland~\cite{mcclelland1986distributed} introduced the back propagation concept.
ANNs are used to estimate a complex function by learning to generalise using the given input values and the corresponding output values. 

An ANN is generally composed of an input layer, zero or more hidden layers, and an output layer. Each layer is composed of one or more neurons. Neurons in layer $i$ are connected to the ones in layer $j$, $j=i+1$. This connection is called weight and is represented as $w_{ij}$. During the training process, the input values are propagated forward, the error is calculated (based on the expected output), then the error is propagated, and the weights are adjusted accordingly. The weight of a connection implies the significance of the input. 

Formally, the output of a single neuron is calculated as shown in equation~\ref{eq:neuron_calc}.

\begin{equation}
    \label{eq:neuron_calc}
    O = f((\sum_{i=0}^{n}{x_i.w_{i}}) + b)
\end{equation}

\noindent
where $n$ represents the number of inputs to this node, $x_i$ is the $i^{th}$ input value, $w_i$ is the weight value, b is a bias value. Finally, $f$ is the activation function, which squashes the output, Activation function can be, but not limited to, Tanh, Sigmoid, and Rectified Linear Unit~(RELU).

The error $E$ is calculated at the final layer using the difference between the expected output and the predicted output (which is, as aforementioned, a result of propagating the input signal). 
Finally, the weights are updated based on Equation~\ref{eq:backprop}

\begin{equation}
    \label{eq:backprop}
    w_{t+1} = w_t - \eta \frac{dE}{dw_t}    
\end{equation}

\noindent where $w_t$ is the old weight and $w_{t+1}$ is the new weight. $\eta$ is the learning rate to control the gradient decent steps. 

The weight of a neuron is directly proportional to the significance of the node's input. This is because the output of any neuron is calculated by multiplying the weights by the input values. 

The next section will discuss the proposed features, then Section~\ref{sec:experiments} outlines the experiments where ANNs are used for the classification purpose. 

\section{Methodology}
\label{sec:methodology}

\begin{figure}[htb]
    \centering
    \includegraphics[width=0.7\textwidth]{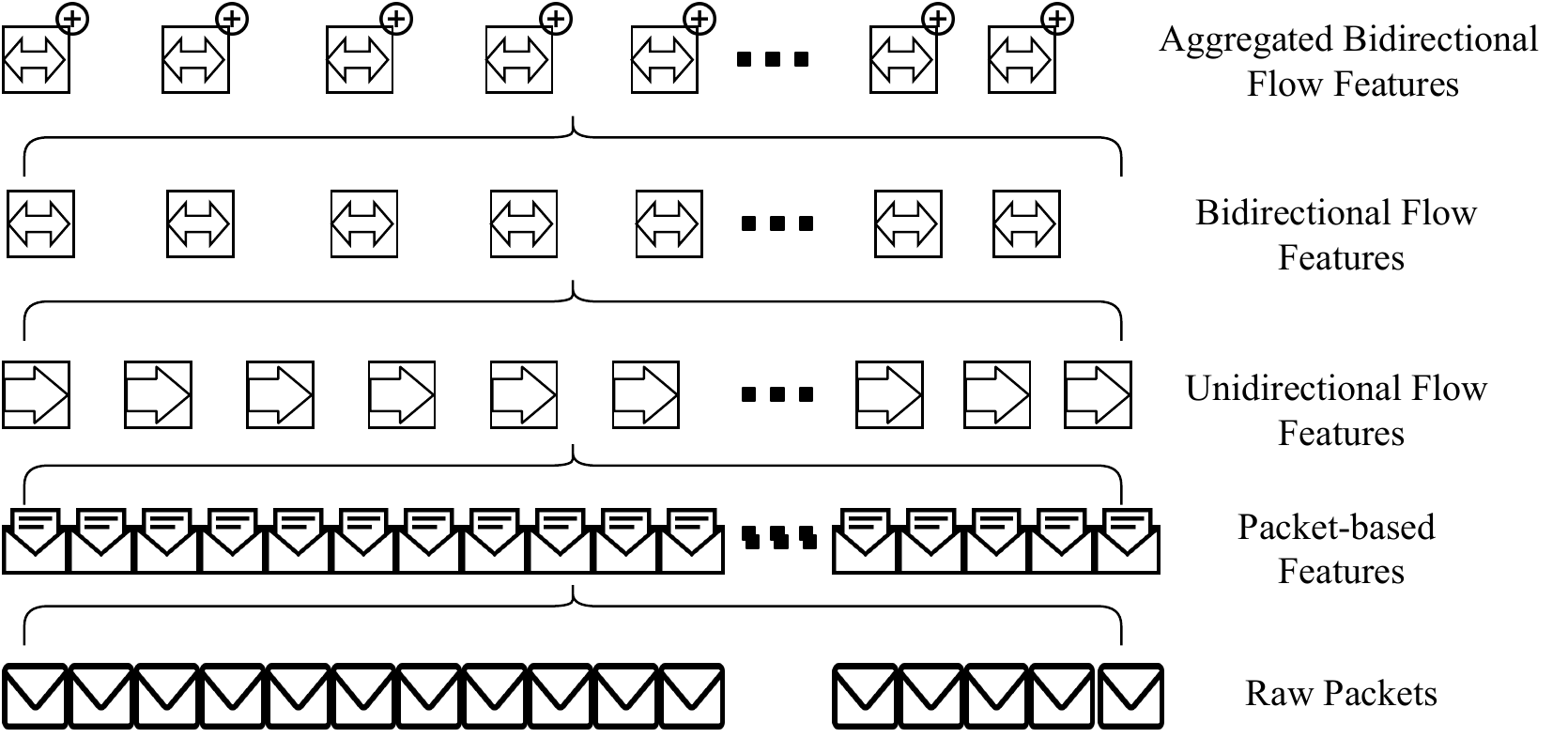}
    \caption{Networking Features Abstraction Levels}
    \label{fig:abstraction-levels}
\end{figure}

Given a raw capture of internet traffic in the form of a raw ``\textit{pcap}'' file, two levels of features are traditionally extracted. As illustrated in Figure~\ref{fig:abstraction-levels}, at the lower level, individual packets are inspected and packet-based features are extracted. These features include flags, packet size, payload data, source and destination address, protocol, etc. At the higher level, flow features (unidirectional and bidirectional) are extracted which consider all the individual packets in a particular communication.

In this paper, a novel additional (third) level of abstraction is proposed where bidirectional flows are grouped into bundles and flow aggregation features are derived. Flow aggregation features aim at representing information about the whole communication between network hosts. These features provide additional traffic characteristics by grouping individual flows. In this setup, it is assumed that legitimate hosts establish secure communication using any of the well-known authentication mechanisms~\cite{s18072394, 9187988}. 
After these aggregated features are calculated, they are propagated back to each bidirectional flow in the bundle/group. This is represented with the superscript + sign in Figure~\ref{fig:abstraction-levels}. The two proposed flow aggregation features in this manuscript are (a)~the number of flows and (b)~the source ports delta. 

The first feature, `Number of flows', represents the number of siblings in a flow bundle. 
Given the communication between a host, $A$, and one or more hosts, all flows initiated by $A$ are counted. This feature represents the communication flow. The advantage of this feature is that it is significant for attacks that intentionally spread their associated requests over time when targeting a single host. However, when grouped, the bundled flow will have additional information about how many flows are in the same group that can resemble the communication pattern. Moreover, it can represent patterns when an attacker targets many hosts, but each with a few communications, when grouped, a pattern can be identified.

\begin{figure}[hbt]
    \centering    \includegraphics{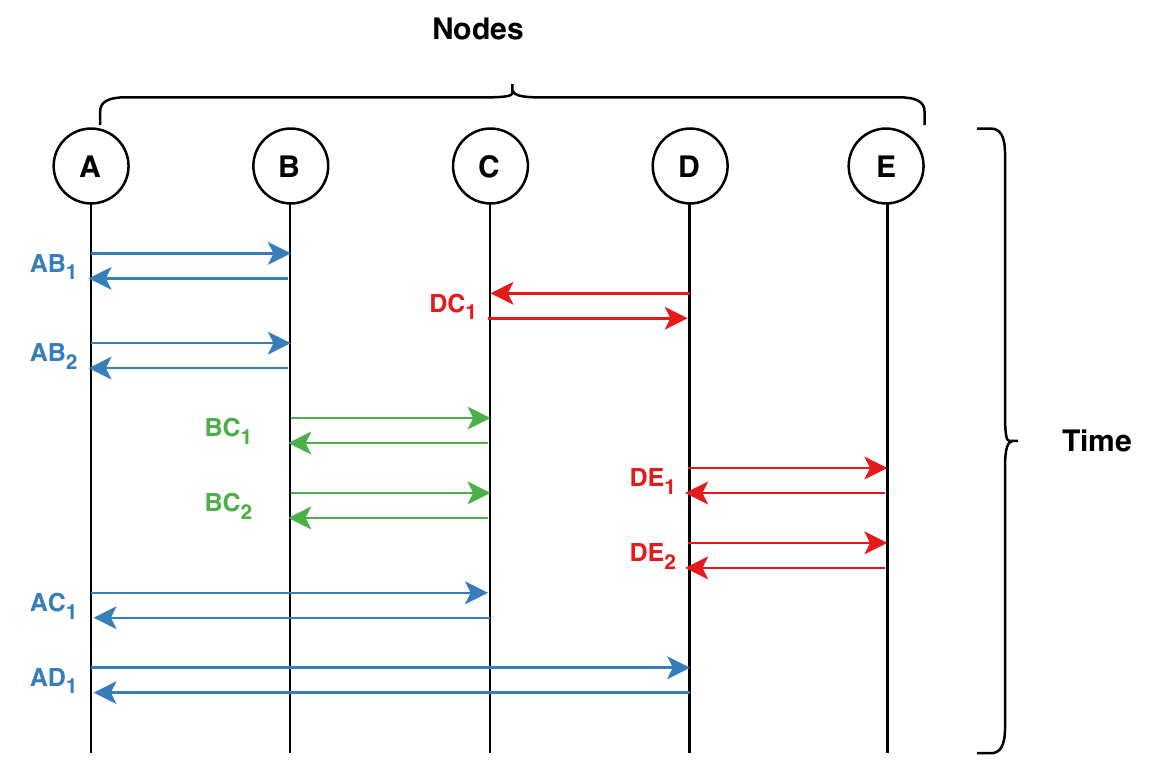}
    \caption{Flow aggregation of network traffic flows. Each colour represents an aggregated flow}
    \label{fig:flowaggregation_timeline}
\end{figure}

Figure~\ref{fig:flowaggregation_timeline} visualises the bundling process of flows. Each letter (Figure~\ref{fig:flowaggregation_timeline} top) represents a node/host in the network. Each double arrow represents a bidirectional flow with the notation $XY_i$, such that $X$ is the source node, $Y$ is the destination node, and $i$ is the communication counter. Finally, the colours in Figure~\ref{fig:flowaggregation_timeline} represent the grouping of flows into bundles. 

With reference to Figure~\ref{fig:flowaggregation_timeline}, the first bundle (blue) has $4$ flows, thus $AB_1$, $AB_2$, $AC_1$, and $AD_1$ will have the `number of flows' feature set to $4$. Similarly, the second bundle (green), $BC_1$ and $BC_2$ will have the value $2$ and so on.

The second feature `source ports delta' demonstrates the ports delta. This feature is calculated using all the port numbers used in a bundle communication flow. Algorithm~\ref{alg:prt_delta-feature} illustrates the feature calculation. The advantage of  this feature is to capture the level and variation pattern of used ports in legitimate traffic. This feature adds this piece of information to each flow, which then enhances the learning and classification as further discussed in Section~\ref{sec:experiments}.

\begin{algorithm}[hbt]
    \caption{Calculate Ports Delta Feature}
    \label{alg:prt_delta-feature}
    \hspace*{\algorithmicindent} \textbf{Input:} List of bundle flow ports \\
    \hspace*{\algorithmicindent} \textbf{Output:} Ports Delta Feature 
    \begin{algorithmic}[1]
    \setlength\baselineskip{17pt}
    \State ports.sort()
    \For{$i \in length(ports) - 1$}
    \State diff[i] $\leftarrow$ abs(ports[i+1] - ports[i])
    \EndFor
    \State avg\_diff $\leftarrow$ diff.mean()
    \State \textbf{return} avg\_diff
    \end{algorithmic}
\end{algorithm}

To validate the significance of the newly added features proposed in this manuscript, a feature selection algorithm is used. 
Recursive Feature Elimination~(RFE)~\cite{rfe} is used to select the best $k$ features to use for classification. 
Over the various experiments discussed in Section~\ref{sec:experiments}, RFE demonstrates that the two features are important for identifying classes that mimic benign behaviour. 
This emphasises the case in which the additional level of feature abstraction is needed. When attackers attempt to mimic benign behaviour, the attacks are in-distinctive using flow features solely. Therefore, the need for new features.

\section{Experiments and Results}
\label{sec:experiments}

In this section, the conducted experiments are explained and the results are discussed. For a dataset to be appropriate to evaluate the proposed features, it has to (a)~include both benign and cyber-attack traffic, (b)~cover benign mimicking cyber-attacks, and (c)~involve multiple attackers to guarantee the aggregation logic. Based on these criteria, 
the CICIDS2017 dataset~\cite{RefWorks:doc:5d4acfdae4b002b95f900284} is selected. The dataset contains a wide range of insider and outsider attacks alongside benign activity. Sarker~\textit{et~al.}~\cite{sarker2020cybersecurity} highlight in their survey that the CICIDS2017 dataset is suitable for evaluating ML-based IDS including zero-day attacks~\cite{sarker2020cybersecurity}. As aforementioned, the focus of this manuscript is on attacks that mimic benign behaviour. 

Therefore, the attacks of interest from the CICIDS2017 dataset are (1)~DoS Slowloris  and (2)~DoS SlowHTTPTest. These two attacks implement low-bandwidth DoS attacks in the application layer. This is done by draining concurrent connections pool~\cite{SlowHTTP10}. Since these two attacks are performed slowly, they are typically hard to detect. 
Alongside the DoS SlowHTTPTest and Slowloris, two other attacks are of interest for comparative purposes; (3)~PortScan and (4)~DoS Hulk. These two attacks resemble the case where attacks are easy to discriminate from benign traffic.

First, each of these four attacks and benign pcap files is processed. The output of this process is bidirectional flow features and aggregation features. 
Second, RFE is used to select the best $k = 5$ features.
Third, the selected features are used as input to an ANN that acts as the classifier. The ANN architecture is composed of 5 input neurons, 1 hidden layer with 3 neurons, and an output layer. 
It is important to mention that since the focus is on the evaluation of the additional level of features and not the classifier complexity, the number of chosen features is small and a straightforward ANN is used. 

Three experiments are performed. The first experiment is a binary classification of each of the attacks of interest (Section~\ref{sec:binary}). The second experiment is a 3-class classification (Section~\ref{sec:three-classes}). This experiment evaluates the classification of benign, a benign-mimicking attack, and a distinctive attack (i.e., do not mimic benign behaviour). Finally, the third experiment is a 5-class classification including all classes of interest (Section~\ref{sec:five-classes}). Each experiment is performed twice, (a)~with the bidirectional features only and (b)~with the bidirectional features and the aggregation features. The RFE is performed independently in each experiment. It is important to highlight that the features selected by RFE confirm the importance of the proposed flow aggregation ones as follows. The top two RFE features for the 5-class classification are `Number of Flows' and `Source ports delta'. For the three class classification of the Slowloris, the top two RFE features are `Source ports delta' and `Number of Flows'. Similarly, the `Number of Flows' feature is chosen by RFE for the 3-class classification and the `Source ports delta' for the binary classification. This confirms the significance of the new features prior to analysing the classification results when using the additional features. 
For the purpose of evaluation  comparison, the RFE features are selected without the flow aggregation ones in consideration.

\subsection{Evaluation Metrics}
In this section, the used evaluation metrics are discussed. The evaluation is performed in a 10-fold cross-validation manner. 
Recall, precision, and F1-score are reported for each experiment. Their formulas are shown in Equation~\ref{eq:recall}, Equation~\ref{eq:precision}, and Equation~\ref{eq:F1}, respectively. True Positive~(TP) represents attack instances correctly classified, False Negative~(FN) represents attack instances misclassified, False Positive~(FP) represents benign instances misclassified. 

\begin{equation}
    \label{eq:recall}
    Recall = \frac{TP}{TP + FN}
\end{equation}

\begin{equation}
    \label{eq:precision}
    Precision = \frac{TP}{TP + FP}
\end{equation}

\begin{equation}
    \label{eq:F1}
   F1 = \frac{2TP}{2TP + FP + FN}
\end{equation}


\vspace{4mm}
\subsection{Binary Classification Results}
\label{sec:binary}
The first experiment is a binary classifier. Each of the attacks of interest is classified against benign behaviour. Table~\ref{tab:slowloris_benign} and Table~\ref{tab:slowhttp_benign} show the precision, recall, and F1-score for DoS Slowloris and DoS SlowHTTPTest, respectively. Moreover, the table lists the five features picked by RFE. By observing the recall of the attack class with the flow aggregation features included, it can be seen that the recall rises from 83.69\% to 91.31\% for the Slowloris attack class and from 65.94\%  to 70.03\% for SlowHTTPTest attack class. Unlike benign mimicking attacks, aggregation features do not provide benefit when classifying attacks that do not mimic benign traffic such as PortScan and DoS Hulk as shown in Table~\ref{tab:hulk_benign} and Table~\ref{tab:portscan_benign}. Precision and recall are high (99\%) for both of these attacks without utilising the aggregation flow features. This is coherent with the discussion in Section~\ref{sec:relatedwork}.

\begin{specialtable}[H]
    \centering
    \caption{Benign vs Slowloris Classification (5-fold cross validation)}

    \begin{tabular}{C{0.16\NetTableWidth}C{0.14\NetTableWidth}C{0.14\NetTableWidth}C{0.14\NetTableWidth}C{0.14\NetTableWidth}C{0.14\NetTableWidth}C{0.14\NetTableWidth}}
    \toprule
        & \multicolumn{3}{c|}{\textbf{Without Aggregation}}& \multicolumn{3}{|c}{\textbf{With Aggregation}} \\ \hline
        \multicolumn{1}{c}{\shortstack[c]{\textbf{RFE} \\ \\ \textbf{Selected} \\ \\  \textbf{Features}}} & \multicolumn{3}{|c}{\shortstack[l]{
            - Fwd Min Inter-arrival Time \\
            - Bwd Min Inter-arrival Time \\
            - Bwd mean time between the first \\packet and each successive packet \\
            - Fwd mean time between the first \\packet and each successive packet \\
            - Fwd STD Inter-arrival Time 
            }}& \multicolumn{3}{|c}{\shortstack[l]{- Without Aggregation Features \\  \qquad+ \\ \textbf{- Number of Flows} \\ \qquad+ \\ \textbf{- Source Ports Delta}}} \\ \midrule

  & \textbf{Precision} & \textbf{Recall} & \textbf{F1} & \textbf{Precision} & \textbf{Recall} & \textbf{F1} \\ \hline 
\textbf{Benign} & $99.04\% \pm 0.08\%$ & $99.86\% \pm 0.13\%$ & $99.45\% \pm 0.05\%$ & $99.49\% \pm 0.08\%$ & $99.99\% \pm 0.01\%$ & $99.74\% \pm 0.04\%$ \\ \hline 
\textbf{Slowloris} & $97.35\% \pm 2.35\%$ & $83.69\% \pm 1.42\%$ & $89.97\% \pm 0.81\%$ & $99.73\% \pm 0.26\%$ & $91.31\% \pm 1.35\%$ & $95.33\% \pm 0.76\%$ \\ \bottomrule 
\end{tabular}
    \label{tab:slowloris_benign}
\end{specialtable}

\begin{specialtable}[H]
    \centering
    \caption{Benign vs SlowHTTPTest Classification (5-fold cross validation)}

    \begin{tabular}{C{0.16\NetTableWidth}C{0.14\NetTableWidth}C{0.14\NetTableWidth}C{0.14\NetTableWidth}C{0.14\NetTableWidth}C{0.14\NetTableWidth}C{0.14\NetTableWidth}}
    \toprule

        & \multicolumn{3}{c|}{\textbf{Without Aggregation}}& \multicolumn{3}{|c}{\textbf{With Aggregation}} \\ \hline
        \multicolumn{1}{c}{\shortstack[c]{\textbf{RFE} \\ \\ \textbf{Selected} \\ \\  \textbf{Features}}} & \multicolumn{3}{|c}{\shortstack[l]{
            - Fwd mean time between the first \\packet and each successive packet \\
            - Bwd mean time between the first \\packet and each successive packet \\
            - Fwd Min Inter-arrival Time \\
            - Bwd Min Inter-arrival Time \\
            - Fwd Max Inter-arrival Time }}& \multicolumn{3}{|c}{\shortstack[l]{- Without Aggregation Features \\  \qquad+ \\ \textbf{- Number of Flows} \\ \qquad+ \\ \textbf{- Source Ports Delta}}} \\ \midrule
  & \textbf{Precision} & \textbf{Recall} & \textbf{F1} & \textbf{Precision} & \textbf{Recall} & \textbf{F1} \\ \hline 
\textbf{Benign} & $98.49\% \pm 0.04\%$ & $99.94\% \pm 0.02\%$ & $99.21\% \pm 0.03\%$ & $98.68\% \pm 0.40\%$ & $99.87\% \pm 0.14\%$ & $99.27\% \pm 0.17\%$ \\ \hline 
\textbf{SlowHTTP Test }& $98.13\% \pm 0.56\%$ & $65.94\% \pm 0.98\%$ & $78.87\% \pm 0.82\%$ & $96.24\% \pm 3.68\%$ & $70.03\% \pm 9.27\%$ & $80.63\% \pm 5.21\%$ \\ \bottomrule 
    \end{tabular}
    \label{tab:slowhttp_benign}
\end{specialtable}

\begin{specialtable}[H]
    \centering
    \caption{Benign vs DoS Hulk Classification (5-fold cross validation)}

    \begin{tabular}{C{0.16\NetTableWidth}C{0.14\NetTableWidth}C{0.14\NetTableWidth}C{0.14\NetTableWidth}C{0.14\NetTableWidth}C{0.14\NetTableWidth}C{0.14\NetTableWidth}}
    \toprule
        & \multicolumn{3}{c|}{\textbf{Without Aggregation}}& \multicolumn{3}{|c}{\textbf{With Aggregation}} \\ \hline
        
        \multicolumn{1}{c}{\shortstack[c]{\textbf{RFE} \\ \\ \textbf{Selected} \\ \\  \textbf{Features}}} & \multicolumn{3}{|c}{\shortstack[l]{
   - Bwd Min Packet Length \\ 
   - Fwd Num Reset Flags \\
   - Bwd Num Push Flags \\
   - Bwd Num Reset Flags \\
   - Fwd Max Inter-arrival Time}}& \multicolumn{3}{|c}{\shortstack[l]{- Without Aggregation Features \\  \qquad+ \\ \textbf{- Number of Flows} \\ \qquad+ \\ \textbf{- Source Ports Delta}}} \\ \midrule

  & \textbf{Precision} & \textbf{Recall} & \textbf{F1} & \textbf{Precision} & \textbf{Recall} & \textbf{F1} \\ \hline 
\textbf{Benign} & $99.83\% \pm 0.04\%$ & $99.99\% \pm 0.01\%$ & $99.91\% \pm 0.02\%$ & $100.00\% \pm 0.00\%$ & $100.00\% \pm 0.00\%$ & $100.00\% \pm 0.00\%$ \\ \hline 
\textbf{Hulk} & $99.98\% \pm 0.03\%$ & $99.51\% \pm 0.10\%$ & $99.74\% \pm 0.06\%$ & $99.99\% \pm 0.02\%$ & $99.99\% \pm 0.02\%$ & $99.99\% \pm 0.02\%$ \\ \bottomrule 
    \end{tabular}
    \label{tab:hulk_benign}
\end{specialtable}

\begin{specialtable}[H]
    \centering
    \caption{Benign vs PortScan Classification (5-fold cross validation)}

    \begin{tabular}{C{0.16\NetTableWidth}C{0.14\NetTableWidth}C{0.14\NetTableWidth}C{0.14\NetTableWidth}C{0.14\NetTableWidth}C{0.14\NetTableWidth}C{0.14\NetTableWidth}}
    \toprule

        & \multicolumn{3}{c|}{\textbf{Without Aggregation}}& \multicolumn{3}{|c}{\textbf{With Aggregation}} \\ \hline
        \multicolumn{1}{c}{\shortstack[c]{\textbf{RFE} \\ \\ \textbf{Selected} \\ \\  \textbf{Features}}} & \multicolumn{3}{|c}{\shortstack[l]{
      - Fwd STD Packet Length \\
      - Bwd Min Packet Length \\
      - Fwd Max Packet Length \\
      - Fwd Mean Packet Length \\
      - Fwd Number of Push Flags }}& \multicolumn{3}{|c}{\shortstack[l]{- Without Aggregation Features \\  \qquad+ \\ \textbf{- Number of Flows} \\ \qquad+ \\ \textbf{- Source Ports Delta}}} \\ \midrule

  & \textbf{Precision} & \textbf{Recall} & \textbf{F1} & \textbf{Precision} & \textbf{Recall} & \textbf{F1} \\ \hline 
\textbf{Benign} & $99.40\% \pm 0.03\%$ & $99.36\% \pm 0.93\%$ & $99.38\% \pm 0.45\%$ & $99.51\% \pm 0.03\%$ & $100.00\% \pm 0.00\%$ & $99.75\% \pm 0.01\%$ \\ \hline 
\textbf{Portscan} & $99.36\% \pm 0.92\%$ & $99.39\% \pm 0.04\%$ & $99.37\% \pm 0.45\%$ & $100.00\% \pm 0.00\%$ & $99.50\% \pm 0.03\%$ & $99.75\% \pm 0.01\%$ \\ \bottomrule  
    \end{tabular}
    \label{tab:portscan_benign}
\end{specialtable}

\begin{figure}[ht!]
    \centering
    \includegraphics[width=0.7\linewidth,trim=2cm 2cm 2cm 2cm,clip=true]{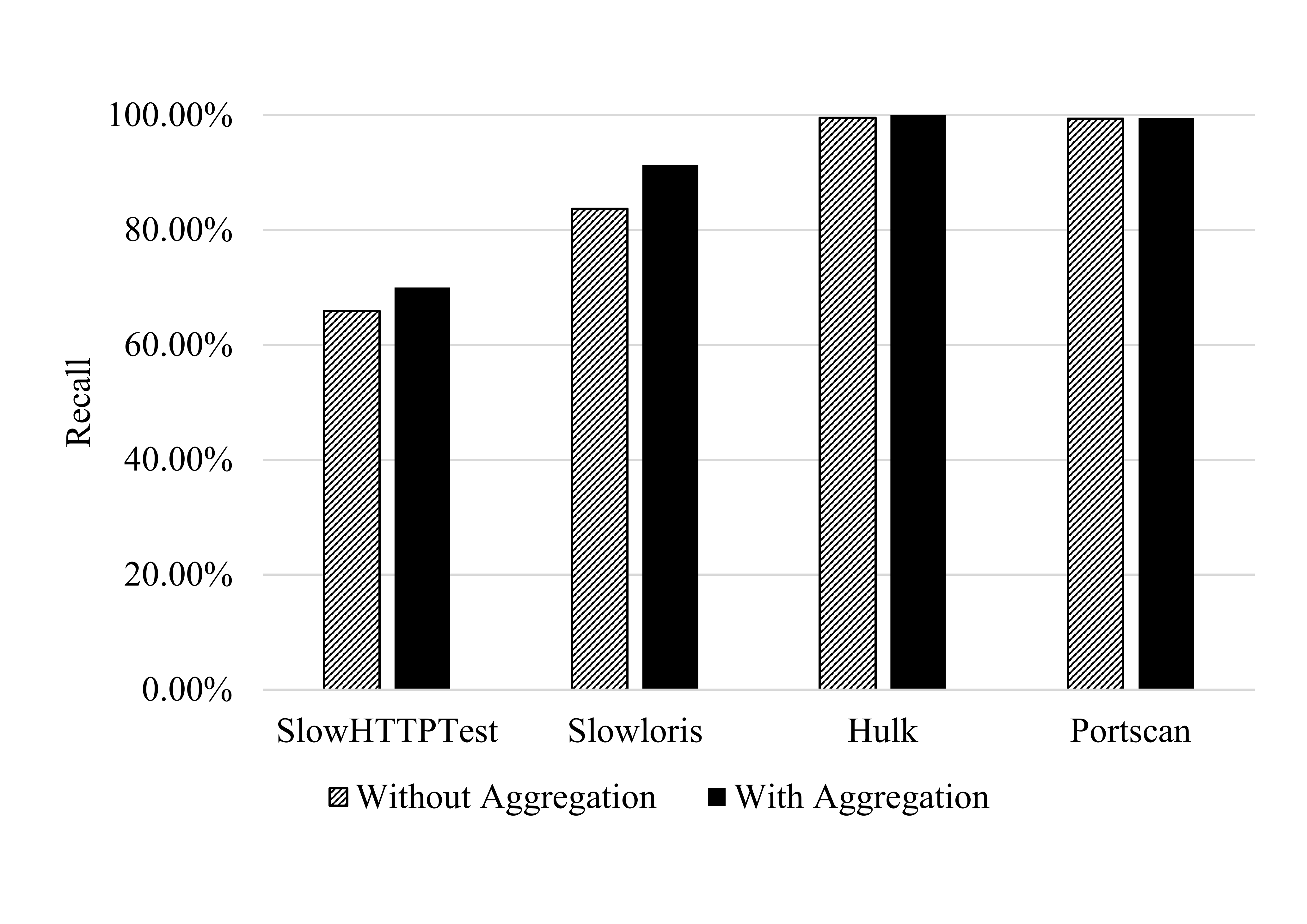}
    \caption{Binary Classification | Impact of Aggregation on Attack Class Recall (Benign vs Attack classification)}
    \label{fig:binary_recall}
\end{figure}

Finally, Figure~\ref{fig:binary_recall} additionally visualises the effect flow aggregation features have on the recall of attack classes. 
It is observed that the two attack classes that mimic benign behaviour (DoS, SlowHTTPTest and Slowloris) have a rise in recall. However, for the other classes (PortScan and DoS Hulk) the aggregation does not impact. This is due to the strength of bidirectional flow features to discriminate classes.

\newpage
\subsection{Three-Classes Classification Results}
\label{sec:three-classes}

In the second experiment, a three-class classification is performed. Benign and PortScan classes (which acts as the discriminative class) are used with each of the other attack classes. 
Similarly, experimental results demonstrate the high recall of both benign and PortScan with and without the use of flow aggregation features and the rise in the other attack recall when flow aggregation features are used. 

By observing Table~\ref{tab:portscan_slowloris_benign}, the recall of DoS Slowloris class rises from 78.25\% to 99.09\% when flow aggregation is used. Similarly, the DoS SlowHTTPTest rises from 0\% to 58.97\% in Table~\ref{tab:portscan_hulk_benign}. Finally, DoS Hulk shows similar behaviour to the binary classification as shown in Table~\ref{tab:portscan_hulk_benign}. 

\begin{specialtable}[H]
    \centering
    \caption{Benign vs PortScan vs  Slowloris Classification (5-fold cross validation)}

        \begin{tabular}{C{0.16\NetTableWidth}C{0.14\NetTableWidth}C{0.14\NetTableWidth}C{0.14\NetTableWidth}C{0.14\NetTableWidth}C{0.14\NetTableWidth}C{0.14\NetTableWidth}}
    \toprule

        & \multicolumn{3}{c|}{\textbf{Without Aggregation}}& \multicolumn{3}{|c}{\textbf{With Aggregation}} \\ \hline
        \multicolumn{1}{c}{\shortstack[c]{\textbf{RFE} \\ \\ \textbf{Selected} \\ \\  \textbf{Features}}} & \multicolumn{3}{|c}{\shortstack[l]{
     - Fwd STD Packet Length \\ 
    - Bwd Min Packet Length \\ 
    - Bwd mean time between the first \\packet and each successive packet \\
    - Fwd mean time between the first \\packet and each successive packet \\
   - Fwd Max Packet Length 
    }}& \multicolumn{3}{|c}{\shortstack[l]{- Without Aggregation Features \\  \qquad+ \\ \textbf{- Number of Flows} \\ \qquad+ \\ \textbf{- Source Ports Delta}}} \\ \midrule

  & \textbf{Precision} & \textbf{Recall} & \textbf{F1} & \textbf{Precision} & \textbf{Recall} & \textbf{F1} \\ \hline 
\textbf{Benign} & $98.31\% \pm 0.16\%$ & $97.69\% \pm 1.02\%$ & $98.00\% \pm 0.49\%$ & $99.46\% \pm 0.04\%$ & $99.99\% \pm 0.01\%$ & $99.73\% \pm 0.02\%$ \\ \hline 
\textbf{Portscan} & $97.85\% \pm 0.99\%$ & $99.60\% \pm 0.12\%$ & $98.71\% \pm 0.45\%$ & $100.00\% \pm 0.01\%$ & $99.50\% \pm 0.03\%$ & $99.74\% \pm 0.01\%$ \\ \hline 
\textbf{Slowloris} & $96.95\% \pm 1.62\%$ & $78.25\% \pm 1.68\%$ & $86.59\% \pm 1.45\%$ & $99.75\% \pm 0.15\%$ & $99.09\% \pm 0.44\%$ & $99.42\% \pm 0.21\%$ \\ \bottomrule 
    \end{tabular}
    \label{tab:portscan_slowloris_benign}
\end{specialtable}

\begin{specialtable}[H]
    \centering
    \caption{Benign vs PortScan vs SlowHTTPTest Classification (5-fold cross validation)}

        \begin{tabular}{C{0.16\NetTableWidth}C{0.14\NetTableWidth}C{0.14\NetTableWidth}C{0.14\NetTableWidth}C{0.14\NetTableWidth}C{0.14\NetTableWidth}C{0.14\NetTableWidth}}
    \toprule

        & \multicolumn{3}{c|}{\textbf{Without Aggregation}}& \multicolumn{3}{|c}{\textbf{With Aggregation}} \\ \hline
        \multicolumn{1}{c}{\shortstack[c]{\textbf{RFE} \\ \\ \textbf{Selected} \\ \\  \textbf{Features}}} & \multicolumn{3}{|c}{\shortstack[l]{
       - Fwd Mean Packet Length \\ 
    - Fwd STD Packet Length \\ 
    - Fwd Max Packet Length \\ 
    - Bwd mean time between the first \\packet and each successive packet \\
    - Fwd mean time between the first \\packet and each successive packet 
      }}& \multicolumn{3}{|c}{\shortstack[l]{- Without Aggregation Features \\  \qquad+ \\ \textbf{- Number of Flows} \\ \qquad+ \\ \textbf{- Source Ports Delta}}} \\ \midrule

  & \textbf{Precision} & \textbf{Recall} & \textbf{F1} & \textbf{Precision} & \textbf{Recall} & \textbf{F1} \\ \hline 
 \textbf{Benign} & $95.10\% \pm 0.04\%$ & $96.40\% \pm 0.12\%$ & $95.75\% \pm 0.06\%$ & $97.67\% \pm 1.25\%$ & $99.99\% \pm 0.01\%$ & $98.81\% \pm 0.64\%$ \\ \hline 
\textbf{Portscan} & $96.45\% \pm 0.11\%$ & $99.52\% \pm 0.05\%$ & $97.96\% \pm 0.06\%$ & $99.99\% \pm 0.02\%$ & $99.42\% \pm 0.15\%$ & $99.70\% \pm 0.08\%$ \\ \hline 
\textbf{SlowHTTP Test} & $0.00\% \pm 0.00\%$ & $0.00\% \pm 0.00\%$ & $0.00\% \pm 0.00\%$ & $79.62\% \pm 39.81\%$ & $58.97\% \pm 29.88\%$ & $67.67\% \pm 34.00\%$ \\ \bottomrule 
    \end{tabular}
    \label{tab:portscan_slowhttp_benign}
\end{specialtable}

\begin{specialtable}[H]
    \centering
    \caption{Benign vs PortScan vs DoS Hulk Classification (5-fold cross validation)}
        \begin{tabular}{C{0.16\NetTableWidth}C{0.14\NetTableWidth}C{0.14\NetTableWidth}C{0.14\NetTableWidth}C{0.14\NetTableWidth}C{0.14\NetTableWidth}C{0.14\NetTableWidth}}
    \toprule

        & \multicolumn{3}{c|}{\textbf{Without Aggregation}}& \multicolumn{3}{|c}{\textbf{With Aggregation}} \\ \hline
        \multicolumn{1}{c}{\shortstack[c]{\textbf{RFE} \\ \\ \textbf{Selected} \\ \\  \textbf{Features}}} & \multicolumn{3}{|c}{\shortstack[l]{
        - Fwd Mean Packet Length \\
        - Fwd Max  Packet Length \\
        - Fwd Number of RST Flags \\
        - Fwd Number of Push Flags \\
        - Bwd Number of RST Flags 
        }}& \multicolumn{3}{|c}{\shortstack[l]{- Without Aggregation Features \\  \qquad+ \\ \textbf{- Number of Flows} \\ \qquad+ \\ \textbf{- Source Ports Delta}}} \\ \midrule

  & \textbf{Precision} & \textbf{Recall} & \textbf{F1} & \textbf{Precision} & \textbf{Recall} & \textbf{F1} \\ \hline 
\textbf{Benign} & $98.10\% \pm 0.05\%$ & $99.30\% \pm 0.95\%$ & $98.69\% \pm 0.49\%$ & $99.57\% \pm 0.25\%$ & $99.94\% \pm 0.04\%$ & $99.75\% \pm 0.12\%$ \\ \hline 
\textbf{Portscan} & $99.10\% \pm 0.94\%$ & $99.39\% \pm 0.04\%$ & $99.24\% \pm 0.48\%$ & $99.95\% \pm 0.03\%$ & $99.73\% \pm 0.24\%$ & $99.84\% \pm 0.11\%$ \\ \hline 
\textbf{Hulk} & $99.94\% \pm 0.03\%$ & $98.56\% \pm 0.06\%$ & $99.25\% \pm 0.03\%$ & $99.98\% \pm 0.03\%$ & $99.50\% \pm 0.06\%$ & $99.74\% \pm 0.03\%$ \\ \bottomrule 
    \end{tabular}
    \label{tab:portscan_hulk_benign}
\end{specialtable}

\begin{figure}[H]
    \centering
    \includegraphics[width=0.7\linewidth,trim=2cm 2cm 2cm 2cm,clip=true]{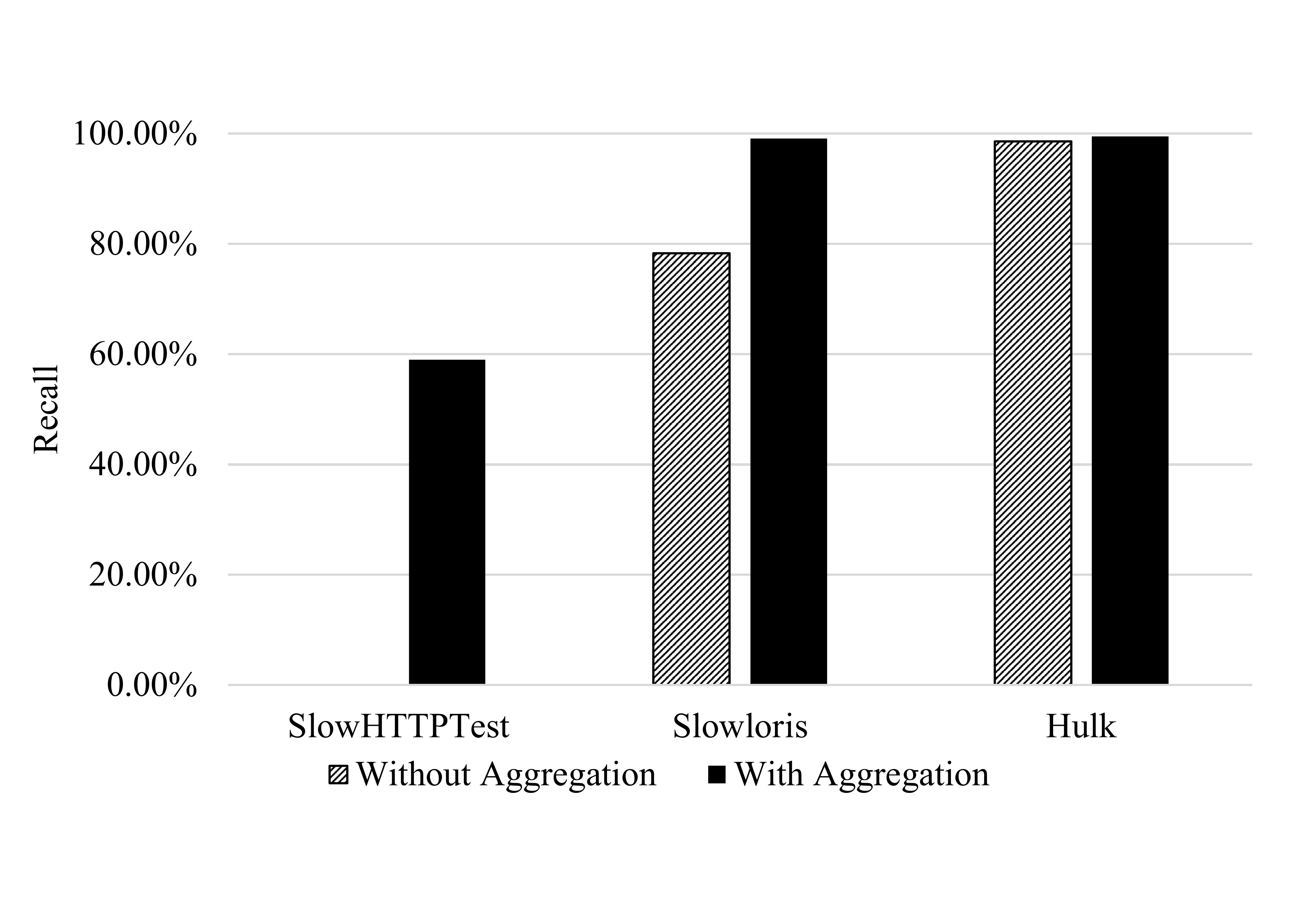}
    \caption{Multi-class Classification | Impact of Aggregation on the Second Attack Class Recall (Benign vs PortScan vs Attack classification)}
    \label{fig:three_recall}
\end{figure}

Figure~\ref{fig:three_recall} visualises the effect flow aggregation features have on the recall of attack classes in a three-class classification problem.
The recall for DoS SlowHTTPTest without using the flow aggregation is 0\%. 

\subsection{Five-Classes Classification Results}
\label{sec:five-classes}

The third experiment combines all the classes of interest in a five-class classification problem. 
Similarly, the experiment is performed twice--- with and without the use of flow aggregation features.
Table~\ref{tab:five_classes_classification_exp1} summarises the performance per class for this experiment. 
A few observations are as follow; (a)~the recall of DoS Slowloris rises from 1.40\% to 67.81\%. (b)~The recall of DoS SlowHTTPTest rises from 0\% to 4.64\% only. This is not because the new features were not significant, but because the model classified DoS SlowHTTPTest as DoS Slowloris. In this case, Flow aggregation features serve to discriminate benign-mimicking attacks from benign traffic but not to discriminate them from each other. Without the aggregation features, 82.84\% of DoS SlowHTTPTest attack instances are classified as benign, however, this drops to 58\% when the flow aggregation features are used.

\begin{specialtable}[H]
    \centering
    \caption{Five Classes Classification (5-fold cross validation)}

     \begin{tabular}{C{0.16\NetTableWidth}C{0.14\NetTableWidth}C{0.14\NetTableWidth}C{0.14\NetTableWidth}C{0.14\NetTableWidth}C{0.14\NetTableWidth}C{0.14\NetTableWidth}}
    \toprule

        & \multicolumn{3}{c|}{\textbf{Without Aggregation}}& \multicolumn{3}{|c}{\textbf{With Aggregation}} \\ \hline
        \multicolumn{1}{c}{\shortstack[c]{\textbf{RFE} \\ \\ \textbf{Selected} \\ \\  \textbf{Features}}} & \multicolumn{3}{|c}{\shortstack[l]{
  - Fwd Mean Packet Length\\
  - Bwd Mean Inter-arrival time \\ 
  - Fwd mean time between the first\\ packet and each successive packet\\ 
  - bwd mean time between the first\\ packet and each successive packet\\
  - Fwd Max packet length
  }}& \multicolumn{3}{|c}{\shortstack[l]{- Without Aggregation Features \\  \qquad+ \\ \textbf{- Number of Flows} \\ \qquad+ \\ \textbf{- Source Ports Delta}}} \\ \midrule

  & \textbf{Precision} & \textbf{Recall} & \textbf{F1} & \textbf{Precision} & \textbf{Recall} & \textbf{F1} \\ \hline 
\textbf{Benign} & $90.97\% \pm 2.99\%$ & $96.77\% \pm 0.80\%$ & $93.74\% \pm 1.33\%$ & $95.11\% \pm 2.11\%$ & $97.11\% \pm 3.30\%$ & $96.05\% \pm 1.84\%$ \\ \hline 
\textbf{Portscan} & $97.12\% \pm 0.80\%$ & $98.90\% \pm 1.05\%$ & $98.00\% \pm 0.38\%$ & $99.92\% \pm 0.13\%$ & $99.41\% \pm 0.15\%$ & $99.67\% \pm 0.08\%$ \\ \hline 
\textbf{Slowloris} & $18.89\% \pm 37.78\%$ & $1.40\% \pm 2.80\%$ & $2.61\% \pm 5.22\%$ & $66.88\% \pm 8.94\%$ & $67.81\% \pm 31.08\%$ & $63.14\% \pm 25.95\%$ \\ \hline 
\textbf{SlowHTTP Test} & $0.00\% \pm 0.00\%$ & $0.00\% \pm 0.00\%$ & $0.00\% \pm 0.00\%$ & $34.12\% \pm 42.81\%$ & $4.64\% \pm 6.51\%$ & $8.10\% \pm 11.21\%$ \\ \hline 
\textbf{Hulk} & $93.75\% \pm 6.47\%$ & $98.61\% \pm 0.73\%$ & $95.98\% \pm 3.14\%$ & $93.00\% \pm 8.58\%$ & $99.34\% \pm 0.15\%$ & $95.85\% \pm 4.87\%$ \\ \hline 
    \end{tabular}
    \label{tab:five_classes_classification_exp1}
\end{specialtable}

To overcome this low recall. Extra features are chosen. Five features are added and the hidden layer neurons are 8. The results of this classification experiment are summarised in Table~\ref{tab:five_classes_classification_exp2}.
The rise in the recall for the attack classes with and without flow aggregation features is as follows; from 33..94\% to 80.39\% and 21.45\% to 64.91\%, for DoS Slowloris and DoS SlowHTTPTest, respectively.

\begin{figure}[H]
    \centering
    \includegraphics[width=0.8\linewidth,trim=1cm 2cm 1cm 2cm,clip=true]{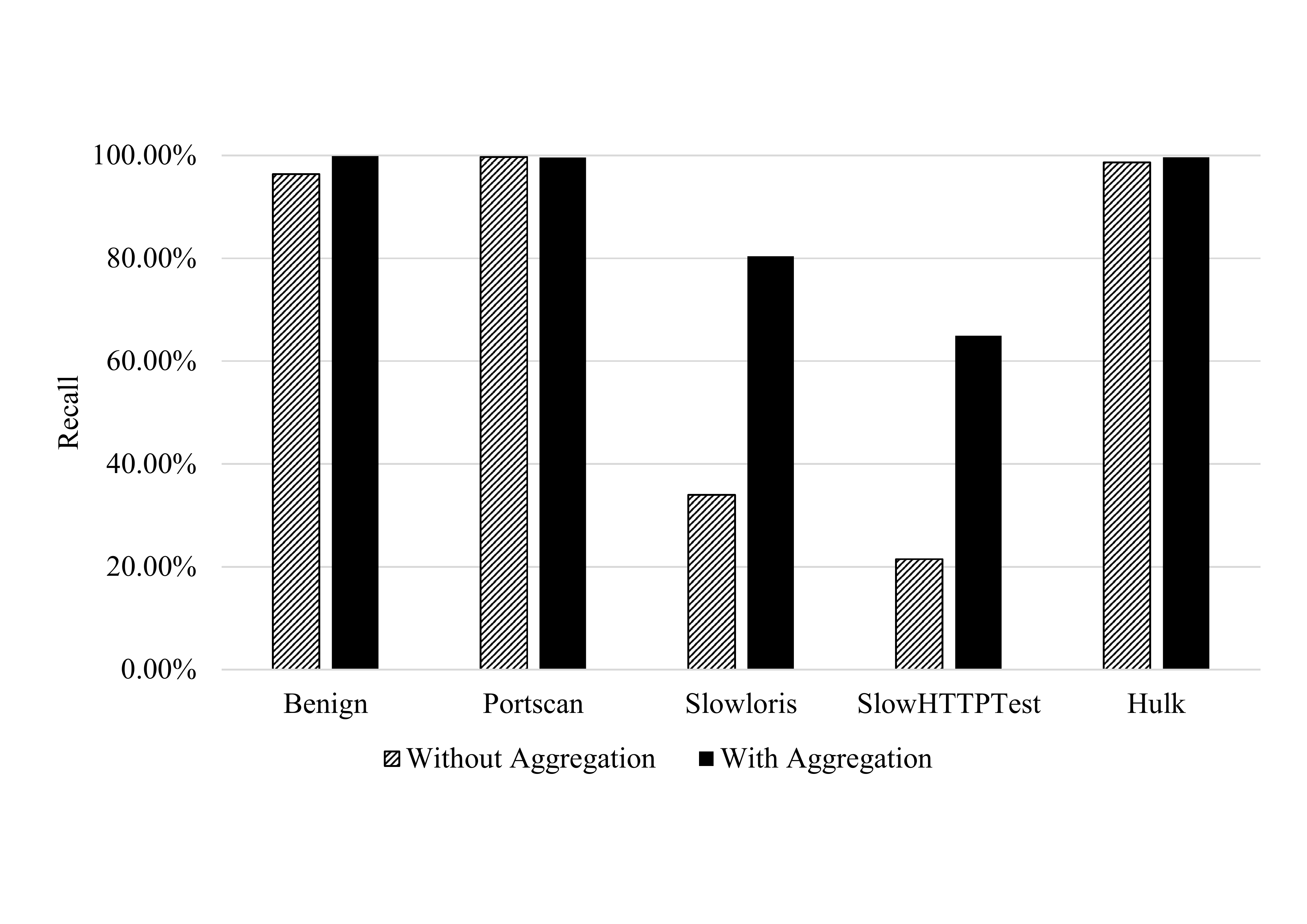}
    \caption{Multi-class Classification | Impact of Aggregation on the Classes Recall)}
    \label{fig:five_recall}
\end{figure}

This behaviour is recognised in Figure~\ref{fig:five_recall}. It is important to mention that there is a rise in the recall of all classes. This rise is more significant for the attack classes that mimic benign behaviour than others. 

\begin{specialtable}[H]
    \centering
    \caption{Five Classes Classification (5-fold cross validation)}

    \begin{tabular}{C{0.16\NetTableWidth}C{0.14\NetTableWidth}C{0.14\NetTableWidth}C{0.14\NetTableWidth}C{0.14\NetTableWidth}C{0.14\NetTableWidth}C{0.14\NetTableWidth}}
    \toprule

        & \multicolumn{3}{c|}{\textbf{Without Aggregation}}& \multicolumn{3}{|c}{\textbf{With Aggregation}} \\ \hline
        \multicolumn{1}{c}{\shortstack[c]{\textbf{RFE} \\ \\ \textbf{Selected} \\ \\  \textbf{Features}}} & \multicolumn{3}{|c}{\shortstack[l]{
    Five RFE features \\  \qquad+ \\
  - Fwd Max Inter-arrival time\\
  - Fwd STD Inter-arrival time\\
  - Fwd Number of Reset Flags\\
  - Fwd Number of Bytes \\
  - Bwd Max Inter-arrival time}}& \multicolumn{3}{|c}{\shortstack[l]{- Without Aggregation Features \\  \qquad+ \\ \textbf{- Number of Flows} \\ \qquad+ \\ \textbf{- Source Ports Delta}}} \\ \midrule

  & \textbf{Precision} & \textbf{Recall} & \textbf{F1} & \textbf{Precision} & \textbf{Recall} & \textbf{F1} \\ \hline 
\textbf{Benign} & $92.37\% \pm 3.56\%$ & $96.34\% \pm 0.11\%$ & $94.28\% \pm 1.83\%$ & $97.35\% \pm 0.53\%$ & $99.90\% \pm 0.13\%$ & $98.61\% \pm 0.31\%$ \\ \hline 
\textbf{Portscan} & $96.48\% \pm 0.07\%$ & $99.74\% \pm 0.03\%$ & $98.08\% \pm 0.04\%$ & $99.84\% \pm 0.18\%$ & $99.59\% \pm 0.20\%$ & $99.71\% \pm 0.10\%$ \\ \hline 
\textbf{Slowloris} & $38.91\% \pm 47.65\%$ & $33.94\% \pm 41.60\%$ & $36.25\% \pm 44.41\%$ & $93.52\% \pm 5.64\%$ & $80.39\% \pm 2.66\%$ & $86.44\% \pm 3.94\%$ \\ \hline 
\textbf{SlowHTTP Test} & $37.52\% \pm 45.98\%$ & $21.45\% \pm 26.27\%$ & $27.29\% \pm 33.44\%$ & $96.80\% \pm 2.72\%$ & $64.91\% \pm 16.75\%$ & $76.61\% \pm 12.25\%$ \\ \hline 
\textbf{Hulk} & $99.92\% \pm 0.14\%$ & $98.63\% \pm 0.67\%$ & $99.27\% \pm 0.29\%$ & $99.88\% \pm 0.14\%$ & $99.69\% \pm 0.21\%$ & $99.78\% \pm 0.15\%$ \\ \hline 
    \end{tabular}
    \label{tab:five_classes_classification_exp2}
\end{specialtable}

\subsection{Zero-Day Attack Detection Revaluation}
The authors' previous work in~\cite{electronics9101684} proposed an autoencoder model to detect zero-day attacks. The model relies on the encoding-decoding capabilities of autoencoders to flag unknown (zero-day) attacks. The model performance is evaluated using all the CICIDS2017 dataset attack classes using three threshold values (0.15, 0.1, and 0.05). The published results demonstrate the ability of the autoencoder to effectively detect zero-day attacks, however, the attacks that mimic benign behaviour experienced very low detection rates. In this section, the published model~\cite{electronics9101684} is reevaluated using the proposed higher level of feature abstraction. The aim is to assess the impact of the new features on zero-day attack detection; specifically benign mimicking ones, whose detection rate is low.

Table~\ref{tab:cicids-ae-results-flow-aggregation} lists the zero-day detection accuracies when flow aggregation features are used alongside the bidirectional flow ones, which were used solely in the previously published work. 
The results show a high detection rate of all attacks, including the attacks that were detected with low accuracy without the flow aggregation features in~\cite{electronics9101684}.

\begin{specialtable}[H]
    \centering
    \caption{CICIDS2017 Autoencoder Zero-day Detection Results using  Flow Aggregation features}
    \label{tab:cicids-ae-results-flow-aggregation}
    \begin{tabular}{C{0.4\NetTableWidth}C{0.2\NetTableWidth}C{0.2\NetTableWidth}C{0.2\NetTableWidth}}
        \toprule
        \textbf{Class} & \multicolumn{3}{c}{\textbf{Accuracy}} \\ \hline
        \textbf{Threshold} & \textbf{0.15 } & \textbf{0.1} & \textbf{0.05} \\ \midrule
        Benign (Validation) & 89.5\% & 85.62\% & 67.59\% \\ \hline
        FTP Brute-force & 99.81\% & 99.92\% & 100\% \\ \hline
        SSH Brute-force & 99.37\% & 100\% & 100\% \\ \hline
        DoS (Slowloris) & 94.12\% & 95.77\% & 100\% \\ \hline
        DoS (GoldenEye) & 100\% & 100\% & 100\% \\ \hline
        DoS (Hulk) & 100\% & 100\% & 100\% \\ \hline
        DoS (SlowHTTPTest) & 99.91\% & 100\% & 100\% \\ \hline
        DDoS & 99.79\% & 100\% & 100\% \\ \hline
        Heartbleed & 99.13\% & 100\% & 100\% \\ \hline
        Web BF & 99.7\% & 99.94\% & 100\% \\ \hline
        Web XSS & 100\% & 100\% & 100\% \\ \hline
        Web SQL & 77.78\% & 77.78\% & 100\% \\ \hline
        Infiltration - Dropbox 1 & 100\% & 100\% & 100\% \\ \hline
        Infiltration - Dropbox 2 & 100\% & 100\% & 100\% \\ \hline
        Infiltration - Dropbox 3 & 46.76\% & 68.68\% & 99.82\% \\ \hline
        Infiltration - Cooldisk & 98.08\% & 100\% & 100\% \\ \hline
        Botnet & 89.83\% & 98.98\% & 100\% \\ \hline
        Portscan & 99.81\% & 99.85\% & 100\% \\ \bottomrule
    \end{tabular}
\end{specialtable}

\begin{figure}[H]
    \centering
    \begin{subfigure}{\linewidth}
        \centering
        \includegraphics[width=\linewidth, trim={1.5cm 5cm 1.5cm 4cm},clip]{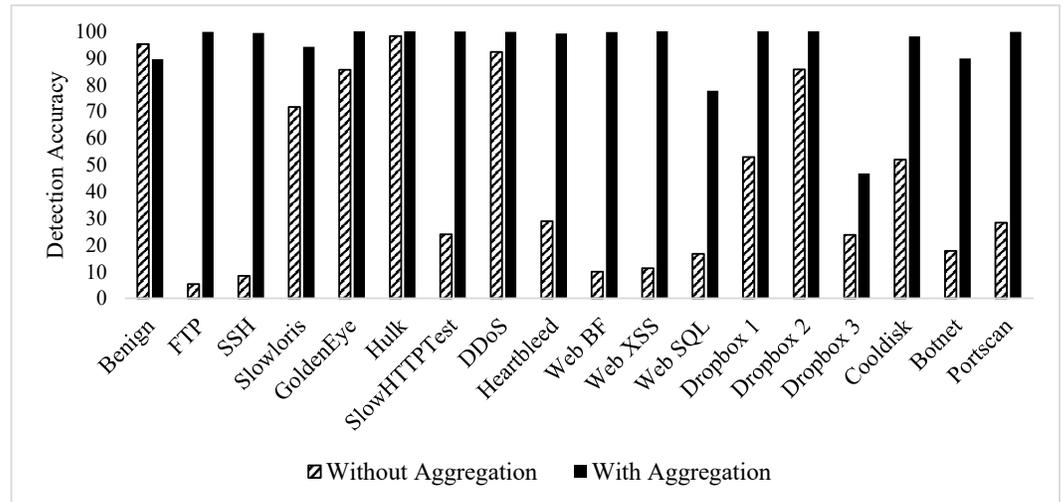}
        \caption{Threshold = 0.15}
    \end{subfigure}%
    
    \begin{subfigure}{\linewidth}
        \centering
        \includegraphics[width=\linewidth, trim={1.5cm 5cm 1.5cm 4cm},clip]{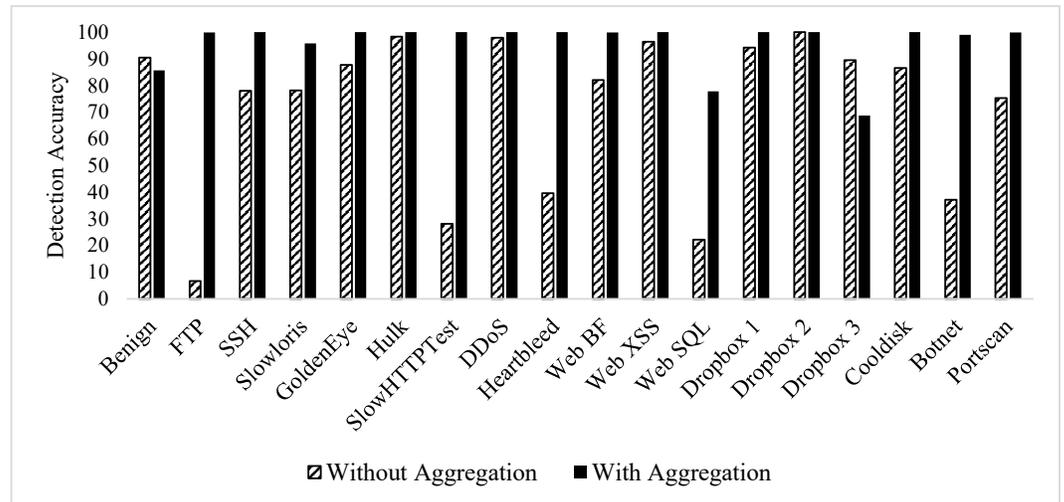}
        \caption{Threshold = 0.1}
    \end{subfigure}
    \caption{CICIDS2017 Autoencoder Zero-Day Detection Comparison Using Flow Aggregation}
    \label{fig:cic-ae-flow-aggregation}
\end{figure}

To visualise the impact of flow aggregation features on zero-day attack detection, Figure~\ref{fig:cic-ae-flow-aggregation} shows the effectiveness of flow aggregation features by contrasting the results that are discussed in this section versus the ones in~\cite{electronics9101684}. It is observed that all attacks experience a rise in detection accuracy when the flow aggregation features are used.

In summary, flow aggregation features prove their effectiveness in providing a deeper representation of a network traffic, thus improving classification performance. This is reflected in the evaluation of both zero-day attacks detection and multi-class attack classifier. The proposed approach has the following three limitations. (a)~The unavailability of traffic flow data affects the flow aggregation features computation. (b)~More features can be derived from the aggregated flows that can further improve the classification accuracy. Lastly, (c)~the proposed features are evaluated using the CICIDS2017, real-time evaluation will provide additional insights. 
\section{Conclusion}
\label{sec:conclusion}

Traditional traffic features have proven powerful when combined with sufficient training examples to train ML-based classifiers and the trained models are capable of classifying some cyber-attacks. However, some cyber-attacks are left undetected. To resolve this limitation, there are two alternatives, (a)~gather huge amounts of data to be able to build even more complex models which is difficult and impractical in some cases and (b)~represent the data using more powerful features. In this paper we appoint the second approach.

This paper presents an additional abstraction level of network flow features. The objective is to improve the cyber-attack classification performance for attacks that mimic benign behaviour. 
Cyber-attacks are becoming more complex and attackers utilise the available knowledge to tailor attacks that can bypass detection tools by acting like benign traffic. 

The idea is to aggregate bidirectional flows to bundles and compute bundle-specific features. Once the features are computed, the values are populated back to the bidirectional flows. The advantage of these additional features is that the bidirectional flows have some additional knowledge/information about their sibling flows.  

The proposal is evaluated using the CICIDS2017 dataset. A group of attacks are used to assess the significance of the new features as well as the performance gain. Four cyber-attack classes are used beside benign class; DoS Slowloris, DoS SlowHTTPTest, DoS Hulk and PortScan. 
ANN is used as the classifier. Three experiments are conducted; binary, three-class and five-class classification. The experiments confirm the need for this additional level of features. The results further demonstrate the significance of the added features for classes that are hard to discriminate from benign, such as DoS Slowloris and DoS SlowHTTPTest. The recall of the cyber-attack classes experiences a high rise when the additional features are used. For example, it is observed that the recall of the DoS Slowloris class rises from 83.97\% with the bi-directional features to 91.31 in binary classification, from 78.28\% to 99.09\% for 3-class classification and from 33.94\% to 80.39\% for 5-class classification.

Furthermore, the additional features prove significant when reassessing the authors' previously published work on detecting zero-day attacks. Benign-mimicking attacks suffered low detection accuracy, however, with the use of flow aggregation features, zero-day attack detection experience a high rise in accuracy.

Future work involves evaluating the significance of the flow aggregation against other operations alongside extracting other high-level features based on needs.

\authorcontributions{Conceptualisation, Hanan Hindy, Robert Atkinson and Xavier Bellekens; 
Formal analysis, Hanan Hindy, Robert Atkinson and Christos Tachtatzis; 
Investigation, Hanan Hindy and Robert Atkinson;
Methodology, Hanan Hindy, Robert Atkinson, Christos Tachtatzis, Miroslav Bures and Xavier Bellekens;
Project administration, Xavier Bellekens; 
Software, Hanan Hindy; 
Supervision, Ethan Bayne and Xavier Bellekens;
Validation, Robert Atkinson, Christos Tachtatzis, Ethan Bayne, Miroslav Bures and Xavier Bellekens;
Writing – original draft, Hanan Hindy; 
Writing – review \& editing, Robert Atkinson, Christos Tachtatzis, Ethan Bayne, Miroslav Bures and Xavier Bellekens.}

\funding{This research received no external funding.}




\dataavailability{The `CICIDS2017' dataset supporting the conclusions of this article is available in the Canadian Institute for Cybersecurity~(CIC) repository, \\ http://www.unb.ca/cic/datasets/ids-2017.html" \\
The code will be available through a GitHub repository.} 


\conflictsofinterest{The authors declare no conflict of interest.} 



\abbreviations{The following abbreviations are used in this manuscript:\\

\noindent 
\begin{tabular}{@{}ll}
DDoS & Distributed Denial of Service\\
DoS & Denial of Service \\ 
DNN & Deep Neural Network \\ 
FN  &False Negative \\ 
FP &False Positive \\ 
IDS & Intrusion Detection System \\ 
RELU & Rectified Linear Unit \\ 
RFE & Recursive Feature Elimination \\ 
ML & Machine Learning \\ 
TP & True Positive
\end{tabular}}

\end{paracol}
\reftitle{References}


\externalbibliography{yes}
\bibliography{7-bibliography.bib}

%


\end{document}